\documentclass[a4paper,11pt]{article}
\pdfoutput=1 % if your are submitting a pdflatex (i.e. if you have
\usepackage{jcappub} % for details on the use of the package, please
\usepackage{color}
\usepackage{amsmath}

\title{\boldmath Quantifying the impact of future Sandage-Loeb test data on dark energy constraints}

\author[a]{Jia-Jia Geng,}
\author[a]{Jing-Fei Zhang,}
\author[a,b,1]{Xin Zhang\note{Corresponding author.}}
\affiliation[a]{Department of Physics, College of Sciences, Northeastern University, \\Shenyang
110004, China}
\affiliation[b]{Center for High Energy Physics, Peking University, \\Beijing 100080, China}
\emailAdd{gengjiajia163@163.com}
\emailAdd{jfzhang@mail.neu.edu.cn}
\emailAdd{zhangxin@mail.neu.edu.cn}

\abstract{
The Sandage-Loeb (SL) test is a unique method to probe dark energy in the ``redshift desert'' of $2\lesssim
z\lesssim 5$, and thus it provides an important supplement to the other dark energy probes.
Therefore, it is of great importance to quantify how the future SL test data impact on the dark energy
constraints.
To avoid the potential inconsistency in data, we use the best-fitting model based on the other
geometric measurements as the fiducial model to produce 30 mock SL test data.
The 10-yr, 20-yr, and 30-yr observations of SL test are analyzed and compared in detail.
We show that compared to the current combined data of type Ia supernovae, baryon acoustic
oscillation, cosmic microwave background, and Hubble constant, the 30-yr observation of
SL test could improve the constraint on $\Omega_m$ by about  80\% and the constraint on
$w$ by about 25\%. Furthermore, the SL test can also improve the measurement of the possible
direct interaction between dark energy and dark matter. We show that the SL test 30-yr data could
improve the constraint on $\gamma$ by about 30\% and 10\% for the $Q=\gamma H\rho_c$
and $Q=\gamma H\rho_{de}$ models, respectively.
}

\begin{document}
\maketitle
\flushbottom
\section{Introduction}
%\label{sec:intro}
Current cosmological observations, such as the type Ia supernovae (SN), the cosmic microwave background (CMB), the baryon acoustic oscillation (BAO),
and so forth, have led to the precision cosmology, in which the geometry and matter/energy contents of the universe are precisely determined.
However, the nature of dark energy still remains enigmatic. To determine the dark energy property, more highly accurate data are needed.
In particular, it is of great importance to combine different observations to precisely determine the equation of state of dark energy.
Different observational data are sensitive in different ways and degrees to dark energy parameters (as well as other parameters), and thus the alone use of
one kind of data would lead to some parameter degeneracy, which might be broken when other kinds of observations are combined.
In the forthcoming years, the close combination of geometric measurements, such as SN and BAO, and structure's growth measurements, such as
clusters of galaxies counts and weak lensing, will definitely play a crucial role in determining the property of dark energy.
On the other hand, some other promising observations will also provide important supplement, significantly affecting the constraints on dark energy.

The Sandage-Loeb (SL) test is a purely geometric measurement of the expansion of the universe, which was firstly proposed by Sandage~\cite{sandage} to
directly measure the temporal variation of the redshift of extra-galactic sources, and then improved by Loeb~\cite{loeb} who provided a
realistic way to measure the redshift drift by detecting the redshift variation of quasar (QSO) Lyman-$\alpha$ absorption lines.
The SL test is rather important due to the fact that it is a unique method to explore the cosmic expansion (and, consequently, dark energy) at the ``redshift desert''
($2\lesssim z\lesssim 5$) that is not covered by any other dark energy probes.
Combining the SL test data from this high-redshift region with other data from low-redshift region, such as SN and BAO, will definitely lead to significant impact
on the dark energy constraints.
The scheduled 39-meter Extremely Large Telescope (ELT) equipped with a high resolution spectrograph called the Cosmic Dynamics Experiment (CODEX) is designed to
perform the SL test. Numerous works in this aspect have been done~\cite{sl1,sl2,sl3,sl4,sl5,sl6,sl7}, some of which assume 240 observed QSOs  in the simulations.
However, according to an extensive Monte Carlo simulation \cite{Liske}, only about 30 QSOs will be bright enough
and/or lie at a high enough redshift for observing redshift drift using a 39-m telescope.
On the other hand, serious and careful quantification of the impact of future SL test data on dark energy constraints based on the observation of 30 QSOs is still absent.
In this paper, we will provide such an analysis.

In existing papers on the SL test, the simulated SL test data are produced based on the fiducial (best-fitting) $\Lambda$CDM model, and constraints on other dark energy models
are performed in the light of these simulated data. Inconsistencies may occur when these simulated SL test data are combined with other actual data, and thus such
a combination may not be appropriate. In order to properly quantify the impact of the future SL test data on dark energy constraints, producing the simulated SL test data
consistent with other actual observations is particularly important and necessary. Here, we note that the SL test alone cannot provide a tight constraint on dark energy owing to the fact that
its data are all within the high-redshift range ($2\lesssim z\lesssim 5$) and the low-redshift data are lacking. Hence, the combination of SL test with low-redshift observations
is a must for constraining dark energy. At present, in order to estimate the potential impact of the future SL test data on dark energy constraints, we have to combine the
simulated SL test data with the currently available actual data covering the low-redshift region. And, to avoid the possible inconsistencies between the actual data and simulated
future data, we will choose the best-fitting dark energy models using the current data as the fiducial models in producing the simulated future data.
Such a scheme is of great importance in effectively quantifying how the future SL test data impact on dark energy constraints.

Typical dark energy models should be chosen as proper examples in the analysis.
In this paper, we focus on the constant $w$ dark energy model. Since dark energy along with cold dark matter (CDM) dominates the universe, we call this model the
$w$CDM model for convenience. In addition, because there is always an important possibility that dark energy directly interacts with cold dark matter, we also consider the interacting
extensions to the $w$CDM model. We call such interacting dark energy models the I$w$CDM models.

In the cases of dark energy interacting with dark matter, the energy balance equations for dark energy and cold dark matter are
\begin{equation}
\label{eqq1}
 \dot{\rho}_{de}+3H\rho_{de}(1+w)=-Q,
\end{equation}
\begin{equation}
\label{eqq2}
 \dot{\rho}_c+3H \rho_c=Q,
\end{equation}
where the Hubble parameter
$H=\dot{a}/a$ describes the expansion rate of the universe and the interacting term $Q$ describes the energy transfer rate between dark energy and dark matter. %densities.
The simplest cases are considered: (i) $Q=\gamma H\rho_c$, with which the model is called the I$w$CDM1 model, and
(ii) $Q=\gamma H\rho_{de}$, with which the model is called the I$w$CDM2 model,
where $\gamma$ is the dimensionless coupling.

\section{Methodology}

For the current observations used in this analysis, we only consider the most typical ones.
We consider a combination of SN, BAO, CMB, and $H_0$, which is the most commonly used data combination in studies of dark energy.
For the SN data, we use the SNLS compilation with a sample of 472 SNe~\cite{snls3}. For the BAO data, we use the $r_s/D_V(z)$ measurements from
6dFGS, SDSS-DR7, SDSS-DR9, and WiggleZ (six data from $z=0.1$ to 0.73, with the inverse of covariance matrix given in Ref.~\cite{wmap9}).
For the CMB data, we use the Planck distance priors given by Ref.~\cite{WW}. Since we investigate the geometric measurements for dark energy constraints,
the use of the CMB distance priors is fairly sufficient. We also use the direct measurement of $H_0$ by the cosmic distance ladder from the HST,
$H_0=73.8 \pm 2.4$ km s$^{-1}$ Mpc$^{-1}$~\citep{Riess2011}.
Dark energy models are firstly constrained by using the SN+BAO+CMB+$H_0$ data combination, and the best-fitting models derived are then chosen as the
fiducial models to produce the simulated SL test data. The resulting simulated SL test data are thus well consistent with the actual current data, and
consequently, the combination of the simulated SL test data with SN, BAO, CMB, and $H_0$ is appropriate.

Of course, in the 6-parameter base $\Lambda$CDM model, the Planck data are in tension with $H_0$ direct measurement at about the 2.5$\sigma$ level, and
are in tension with the SNLS data at about the 2$\sigma$ level~\cite{Planck}. But, it is shown that the tension on $H_0$ could be greatly relieved once the dynamical dark energy is considered~\cite{hde}
(or a sterile neutrino species is included in the model~\cite{zx14,WHu14,Zhang:2014nta,Li:2014dja}). The tension of Planck with SN may be due to some unknown systematic error sources in SN data, such as the absence of the consideration of
possible evolution of the color-luminosity parameter $\beta$, as analyzed in Refs.~\cite{beta1,beta2}. However, in this paper, we only focus on the impact of the SL test data, and so
we omit the tiny inconsistency between Planck and SN.

As mentioned above, the SL test is a way to measure the redshift variation of quasar Lyman-$\alpha$ absorption lines.
The velocity shift can be detected by subtracting the spectral templates of a quasar taken at two different times, and the quasar systems are now readily targeted and observable in the redshift range $2\lesssim z\lesssim 5$.
%Consider an isotropic source emitting at rest without peculiar motion
The redshift variation is defined as a spectroscopic velocity shift and expressed as~\cite{loeb}
\begin{equation}\label{eq1}
\ \Delta v \equiv \frac{\Delta z}{1+z}=H_0\Delta t_o\bigg[1-\frac{E(z)}{1+z}\bigg],
\end{equation}
where $E(z)=H(z)/H_0$ and $\Delta t_o$ is the time interval of observation.
%{\color{red}The full expression for $E(z)$ is presented in Sec.~\ref{sec:appendix} for $w$CDM, I$w$CDM1, and I$w$CDM2 model, respectively.}

For the dark energy models considered in this paper, namely the $w$CDM, the I$w$CDM1, and the I$w$CDM2 models, the corresponding expressions of $E(z)$ 
are explicitly given in the following,
\begin{equation}
\label{eq_q0}
\begin{split}
 E(z)^2 =\Omega_{r} (1+z)^4 + \Omega_{b} (1+z)^3 + \Omega_{c} (1+z)^3 \\
 +(1-\Omega_{r}-\Omega_{b}-\Omega_{c}) (1+z)^{3(1+w)},
 \end{split}
\end{equation}
\begin{equation}
\label{eq_q1}
\begin{split}
E(z)^2 = \Omega_{r} (1+z)^4 + \Omega_{b} (1+z)^3 + \frac{3w}{\gamma+3w}\Omega_{c} (1+z)^{3-\gamma}  \\
+ (1-\Omega_{r}-\Omega_{b}-\frac{3w}{\gamma+3w}\Omega_{c})(1+z)^{3(1+w)},
\end{split}
\end{equation}
\begin{equation}
\label{eq_q2}
\begin{split}
 E(z)^2 = \Omega_{r} (1+z)^4 + \Omega_{b} (1+z)^3 + \frac{3w\Omega_{c}+\gamma(1-\Omega_{r}-\Omega_{b})}{\gamma+3w} (1+z)^3 \\
 + \frac{3w(1-\Omega_{r}-\Omega_{b}-\Omega_{c})}{\gamma+3w}(1+z)^{3(1+w)+\gamma},
 \end{split}
\end{equation}
where $\Omega_r$, $\Omega_b$, and $\Omega_c$ are the present-day density parameters for the radiation, baryons, and cold dark matter, respectively, 
and $\Omega_r=\Omega_m/(1+z_{\rm eq})$, with $\Omega_m=\Omega_c+\Omega_b$ and $z_{\rm eq}=2.5\times 10^4\Omega_m h^2(T_{\rm cmb}/2.7{\rm K})^{-4}$ 
(here we take $T_{\rm cmb}=2.7255$ K).

Using Monte Carlo simulations and considering not just the Ly$\alpha$ forest but all available absorption
lines, including metal lines, over the entire accessible optical
wavelength range down to a QSO's Ly$\gamma$ emission line, the accuracy of the
spectroscopic velocity shift measurements expected by CODEX can be
expressed as~\cite{Liske}
\begin{equation}\label{eq2}
\sigma_{\Delta v}=1.35
\bigg(\frac{S/N}{2370}\bigg)^{-1}\bigg(\frac{N_{\mathrm{QSO}}}{30}\bigg)^{-1/2}\bigg(
\frac{1+z_{\mathrm{QSO}}}{5}\bigg)^{f}~\mathrm{cm}~\mathrm{s}^{-1},
\end{equation}
where $S/N$ is the signal-to-noise ratio defined per 0.0125 $\mathring{A}$ pixel, $N_{\mathrm{QSO}}$ is the
number of observed quasars, $z_{\mathrm{QSO}}$ represents their
redshifts, and the exponent $f=-1.7$ for $2<z<4$ and $f=-0.9$ for $z>4$. We choose $N_{\mathrm{QSO}}=30$ mock data uniformly distributed among the six redshift bins of $z_{\mathrm{QSO}}\in [2,~5]$. The parameters in Eq.~(\ref{eq1}) are taken to be the best-fit values in the fit to the SN+BAO+CMB+$H_0$ data, and error bars are computed from Eq.~(\ref{eq2})
with a $S/N=3000$. We perform an MCMC likelihood analysis~\cite{cosmomc}
to obtain ${\cal O}$($10^6$) samples for each model considered in this paper.

\section{Results and discussions}

%\begin{table*}\tiny
%\caption{Fit results for the $w$CDM, I$w$CDM1, and I$w$CDM2 models using the
%current only and current+SL 30-yr data.}
%\label{table1}
%\begin{tabular}{cccccccccc}
%\\
%\hline\hline &\multicolumn{3}{c}{current only} &&\multicolumn{3}{c}{current + SL 30-yr} \\
%           \cline{2-4}\cline{6-8}
%Parameter  & $w$CDM & I$w$CDM1 & I$w$CDM2 & & $w$CDM & I$w$CDM1 & I$w$CDM2\\ \hline
%
%
%$\Omega_{m}$       & $0.2844^{+0.0104}_{-0.0093}$
%                   & $0.2849^{+0.0114}_{-0.0085}$
%                   & $0.2834^{+0.0086}_{-0.0114}$&
%                   & $0.2846^{+0.0017}_{-0.0020}$
%                   & $0.2847^{+0.0032}_{-0.0031}$
%                   & $0.2840^{+0.0035}_{-0.0049}$\\
%
%$H_0$              & $70.74^{+1.26}_{-1.30}$
%                   & $72.81^{+1.87}_{-1.67}$
%                   & $71.09^{+1.39}_{-1.10}$&
%                   & $70.71^{+0.45}_{-0.45}$
%                   & $72.95^{+1.23}_{-1.33}$
%                   & $71.01^{+0.77}_{-0.63}$\\
%
%$w$                & $-1.103^{+0.058}_{-0.058}$
%                   & $-1.136^{+0.062}_{-0.055}$
%                   & $-1.152^{+0.064}_{-0.072}$&
%                   & $-1.104^{+0.044}_{-0.043}$
%                   & $-1.134^{+0.047}_{-0.046}$
%                   & $-1.148^{+0.052}_{-0.064}$\\
%
%$\gamma$           & ...
%                   & $-0.0112^{+0.0054}_{-0.0078}$
%                   & $-0.0284^{+0.0218}_{-0.0206}$&
%                   & $...$
%                   & $-0.0118^{+0.0050}_{-0.0043}$
%                   & $-0.0256^{+0.0156}_{-0.0224}$\\
%
%\hline
%\end{tabular}
%\end{table*}

\begin{figure}
\begin{center}
\includegraphics[width=12cm]{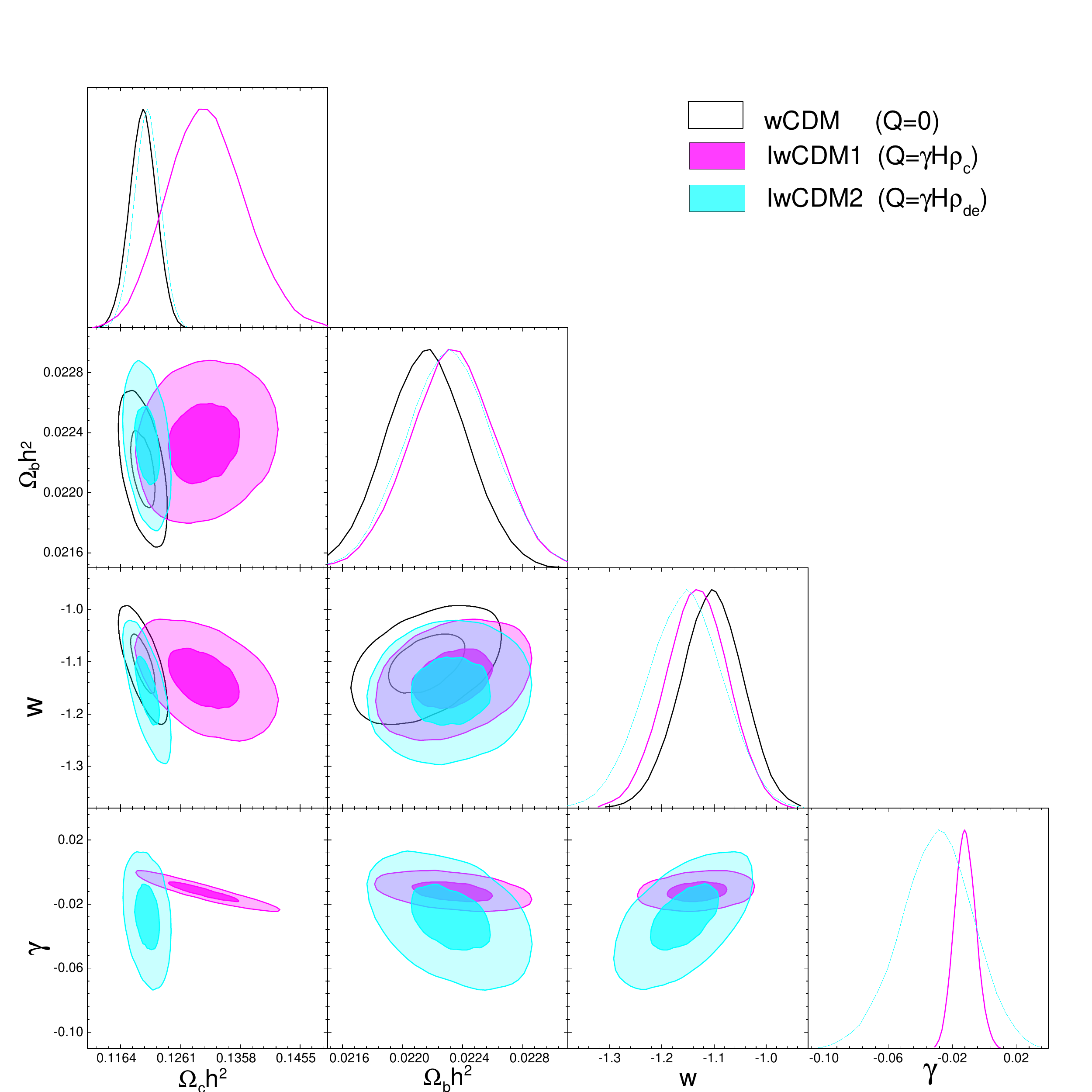}
\end{center}
\caption{Constraints on the $w$CDM, I$w$CDM1, and I$w$CDM2 models from the
current SN+BAO+CMB+$H_0$ data.}
\label{fig1}
\end{figure}

Constraints on the $w$CDM (black), I$w$CDM1 (purple), and I$w$CDM2 (cyan) models from the current
SN+BAO+CMB+$H_0$ data are shown in Fig.~\ref{fig1}.
%Detailed fit values are given in Table~\ref{table1}.
From the figure, one can clearly see that for all the three models $w<-1$ is preferred at more than 2$\sigma$ level.
%For the $w$CDM and I$w$CDM2 models, $\Omega_c h^2$ can be tightly constrained, and a smaller value is
%preferred. But for the I$w$CDM1 model, $\Omega_c h^2$ cannot be well constrained, and a bigger value is more
%favored in this case.
%On the contrary,
The coupling $\gamma$ is tightly constrained in the I$w$CDM1 model, but its constraint is
much weaker in the I$w$CDM2 model.
In both cases, $\gamma<0$ is slightly favored, deviating from the noninteracting case ($\gamma=0$) at about the
1.4$\sigma$ level. (A $\gamma<0$ indicates the energy transfer from dark matter to dark energy, in our convention.)

\begin{figure}
\begin{center}
\includegraphics[width=12cm]{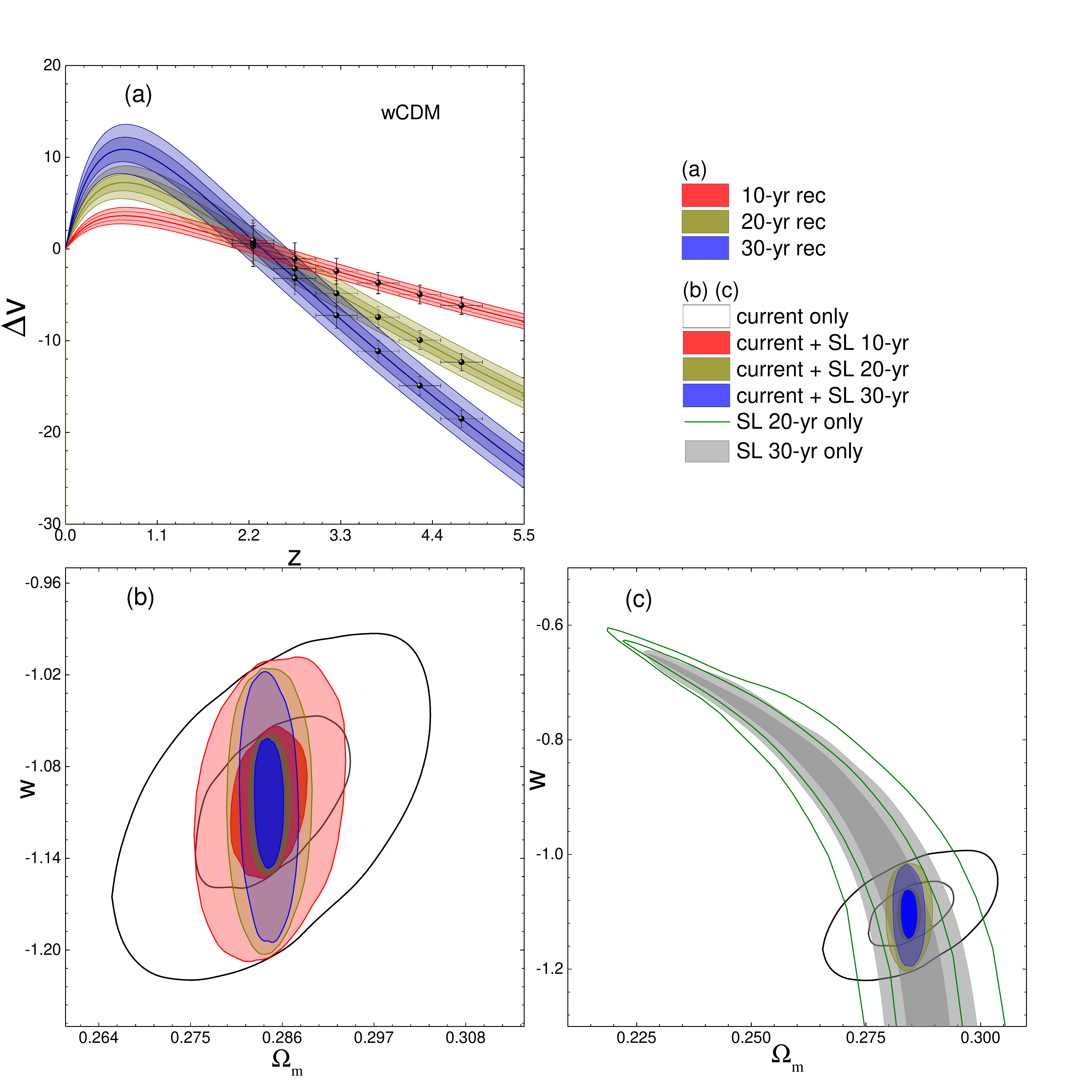}
\end{center}
\caption{The SL test for the $w$CDM model. (a) Reconstructed velocity drift using the current
SN+BAO+CMB+$H_0$ constraint results for 10-yr, 20-yr, and 30-yr observations of SL test 
(red, green, and blue bands).
Comparison is made with the error bars of the simulated SL test mock data.
(b) Combined constraints on the $w$CDM model with current only, current+SL 10-yr, current+SL 20-yr,
and current+SL 30-yr data.
(c) Comparison of the constraints in the $\Omega_m$--$w$ plane from current only and SL test only data.
Note that the orientations of the degeneracies of the two are nearly orthogonal,
so the SL test is highly efficient in breaking the degeneracy in the
other geometric measurements. 
In the SL test simulations, we assume $S/N=3000$ and $N_{\rm{QSO}}=30$, where 
$S/N$ is the signal-to-noise ratio defined per 0.0125 $\mathring{A}$ pixel and 
$N_{\mathrm{QSO}}$ is the number of observed quasars, and we 
also assume that the 30 mock data are uniformly distributed among the six redshift bins of 
$z_{\mathrm{QSO}}\in [2,~5]$.}
\label{fig2}
\end{figure}

Simulating the 30 mock SL data using the best-fitting model as the fiducial model, and then combining them with
the current real data, one can perform a joint constraint on the chosen dark energy model.
The $w$CDM model is taken as a concrete example, and a detailed analysis is made for it.
Figure~\ref{fig2} summarizes the results of our analysis.

In order to directly compare the accuracies of the current data and the future SL data, we reconstruct the
velocity shifts for the $w$CDM model in the redshift range from $z=0$ to 5.5 by using the fit results from the
current SN+BAO+CMB+$H_0$ data, plotted in Fig.~\ref{fig2}(a) as colored bands.
Red, green, and blue bands are for the 10-yr, 20-yr, and 30-yr velocity-shift reconstructions, respectively.
The error bars in the SL test, estimated by Eq.~(\ref{eq2}), are also plotted in Fig.~\ref{fig2}(a) on the
corresponding bands, for a direct comparison.
One can see that the 10-yr observation of SL test is evidently worse than the current combined observations
in accuracy, but a 20-yr observation can greatly improve the accuracy. The 30-yr observation of SL test
gets further improvement, basically comparable with the current combined data in accuracy.
So, the SL test, as a high-redshift supplement of $2\lesssim z\lesssim 5$ to other cosmological geometric
measurements, will play an important role in probing the expansion history of the universe and thus the
property of dark energy.

Figure~\ref{fig2}(b) shows a comparison of the joint constraints on the $w$CDM model in the $\Omega_m$--$w$
plane. The 68\% and 95\% CL posterior distribution contours are shown, where the current only,
the current+SL 10-yr, the current+SL 20-yr, and the current+SL 30-yr results are colored with white, red, green,
and blue, respectively. It is clear that using only the SL 10-yr combined data can improve the constraint significantly.
Of particular importance is that the combination with the SL test can tightly constrain $\Omega_m$ to a
high precision. With an only 10-yr observation, the constraint of $\Omega_m$ will be improved by 51.3\%.
That is an enormous improvement.
%Note that we refer to the 1$\sigma$ improvement when the comparison of different observations is mentioned.
The constraint on $w$ can also be improved significantly. With an only
10-yr observation, the constraint of $w$ is improved by 12.1\%, which is also remarkable.
With the 20-yr and 30-yr observations, the constraints on $\Omega_m$ are improved by 72.8\% and 81.2\%,
respectively, and the constraints on $w$ are improved by 18.1\% and 25.0\%, respectively.
Therefore, we can conclude that using a 30-yr observation of the SL test the geometric constraints on dark energy
would be improved enormously, i.e., the constraints on $\Omega_m$ and $w$ would be improved,
relative to the current joint observations, by at least 80\% and 25\%, respectively.
Note that this is a conservative estimation since 30 years later other geometric measurements themselves
would also be improved greatly in accuracy, and the combination of these highly accurate data could
constrain dark energy more tightly.

In Fig.~\ref{fig2}(c), we show how the SL test breaks the degeneracy in the current data constraints.
Solely using the SL test data cannot place tight constraints on dark energy due to the lack of low-redshift
data. In this case, strong degeneracy between $w$ and $\Omega_m$ will appear, as shown in Fig.~\ref{fig2}(c)
where the 20-yr and 30-yr observations are taken as examples.
However, one can also directly see that the orientations of the degeneracies for the SL test and the current
combined data are nearly orthogonal, and so the SL test is highly efficient in breaking the degeneracy in the
other geometric measurements.
The 68\% and 95\% CL contours from the current+SL 20-yr (green) and current+SL 30-yr (blue)
combined constraints are also plotted, from which one can directly see the crucial role played by the SL test
in breaking the degeneracy.
Further comparing the 20-yr and 30-yr observations of the SL test, we find that the 30-yr observation improves,
relative to the 20-yr one, the constraints on $\Omega_m$ and $w$ by 30.2\% and 8.4\%, respectively.
Therefore, in the following, to show the decisional impact on dark energy constraints from the SL test, we
use the 30-yr observation as an example.

\begin{figure}
\begin{center}
\includegraphics[width=12cm]{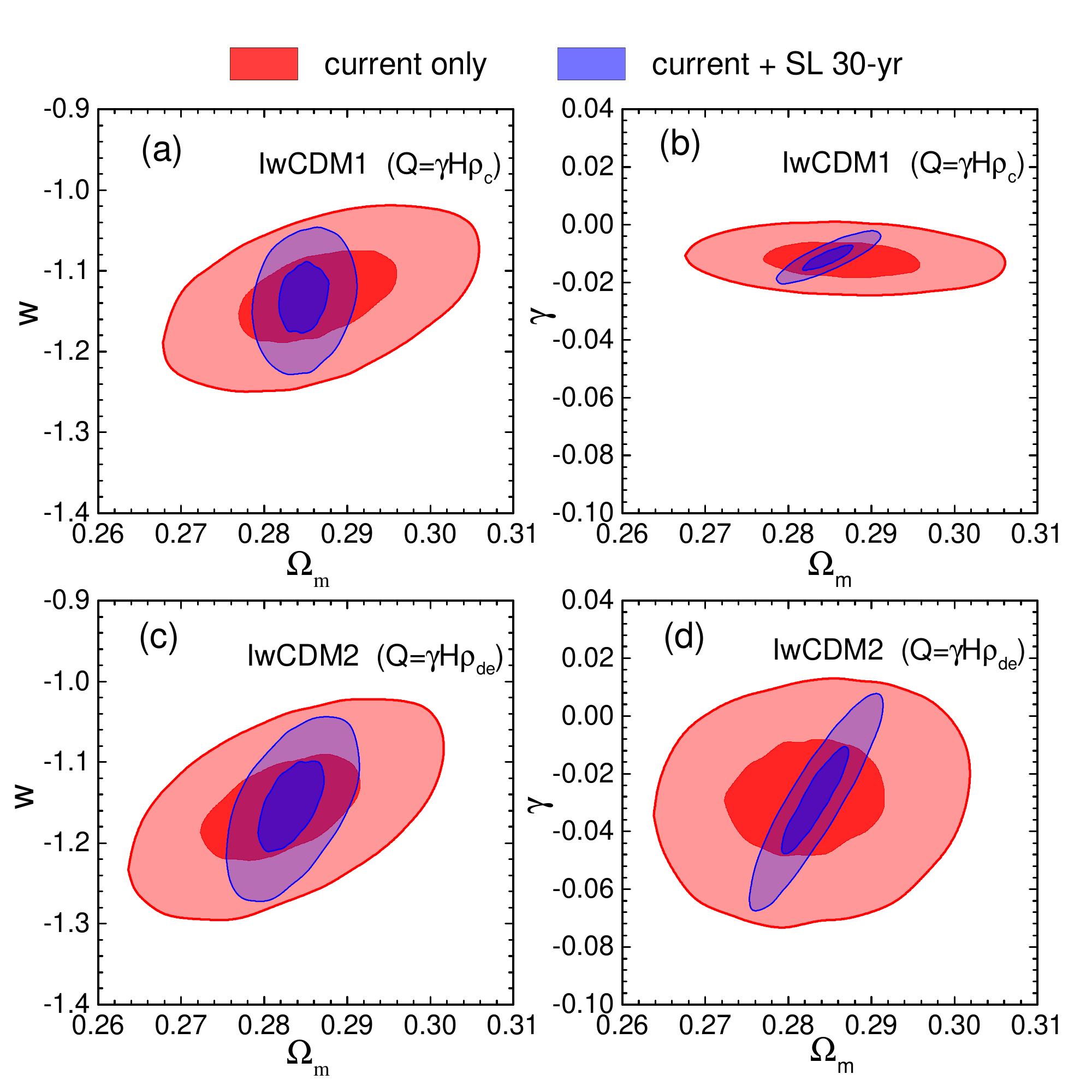}
\end{center}
\caption{Constraints on the two I$w$CDM models from the current only and
current+SL 30-yr data.   }
\label{fig3}
\end{figure}

At last, we discuss how the SL test helps improve the measurement of the possible direct interaction between
dark energy and dark matter.
%In Fig.~\ref{fig3}, we show the joint constraints on the I$w$CDM1 and I$w$CDM2
%models; the constraints on the $w$CDM are also shown for a comparison.
%(Detailed fit results can be found in Table~\ref{table1}.)
%In the left three panels, we show the reconstructed velocity shift of 30-yr observation from the current combined
%SN+BAO+CMB+$H_0$ constraints, and the error bars from the SL 30-yr observation are also shown for a direct
%comparison. In the middle three panels, we show the combined constraints on the three models in the
%$\Omega_m$--$w$ plane. In the right two panels, we show the combined constraints on the two I$w$CDM models
%in the $\Omega_m$--$\gamma$ plane.
In Fig.~\ref{fig3}, we show the joint constraints on the two I$w$CDM models in the
$\Omega_m$--$w$ and the $\Omega_m$--$\gamma$ planes.
The impact of the SL test can be directly seen from the comparison of the red (current only)
and blue (current+SL 30-yr) contours.
Compared to the current only data, the current+SL 30-yr data improve the constraints on $\Omega_m$,
$w$, and $\gamma$ by 68.3\%, 20.5\%, and 29.5\%, respectively, for the I$w$CDM1 model, and by
58.0\%, 14.7\%, and 10.4\%, respectively, for the I$w$CDM2 model.
Therefore, we find that as a purely geometric measurement, the SL test can also mildly improve
the constraints on the interaction between dark energy and dark matter.

\section{Summary}
%\label{sec:concl}
In this paper, we have discussed the quantification of the impact of future SL test data on
dark energy constraints. The SL test is the unique dark energy probe
of the ``redshift desert'' of $2\lesssim z\lesssim 5$ and so it is particularly important,
as a purely geometric measurement,  for constraining
dark energy as a supplement to other important geometric measurements.
To quantify its impact, we should guarantee that the simulated future SL test data are
consistent with the other geometric measurement data. Thus, we used the best-fitting models based on the
other combined geometric measurement data as the fiducial models to produce the mock SL test data,
and then do the analysis with these data.
Typical dark energy models, such as the $w$CDM model and its interacting extensions (I$w$CDM models),
were considered in this analysis. Since we only focus on the purely geometric measurements,
we do not consider the cosmological perturbations in the dark energy models.
But we also note here that the cosmological perturbations would always be stable in all the interacting
dark energy models, provided that a new framework for calculating the perturbations is adopted, as
analyzed in detail in Ref.~\cite{Li:2014eha}.
For related discussions concerning the stability, see also Ref.~\cite{Li:2013bya}.

With the help of the SL test data, the constraints on dark energy will be improved significantly,
compared to the current SN+BAO+CMB+$H_0$ constraint results.
The 10-yr, 20-yr, and 30-yr observations of SL test have been analyzed and compared in detail, and we
found that the 30-yr observation could provide fairly important supplement to the other observations.
The 30-yr observation of SL test is able to improve, relative to the current combined data, the constraint of
$\Omega_m$ by about 80\% and the constraint of $w$ by about 25\%, according to our analysis.
In addition, for the I$w$CDM model with $Q=\gamma H\rho_c$, the constraint on $\gamma$ is improved by
about 30\%, and for the I$w$CDM model with $Q=\gamma H\rho_{de}$, the constraint on $\gamma$ is
improved by about 10\%. Therefore, according to the quantification study, the SL test is proven to be a
very important supplement to the other geometric measurement observations, and will definitely play a
crucial role in constraining dark energy in the future.

\acknowledgments

We acknowledge the use of {\tt CosmoMC}. We thank Yun-He Li for helpful discussion.
J.F.Z. is supported by the Provincial Department of Education of
Liaoning under Grant No. L2012087. X.Z. is supported by the National Natural Science Foundation of
China under Grant No. 11175042 and the Fundamental Funds for the 
Central Universities under Grant No. N120505003.

\end{document}